\documentclass{WileyMSP-template}
\usepackage{amsmath,amsfonts,amssymb,xcolor}

\begin{document}

\pagestyle{fancy}
\rhead{\includegraphics[width=2.5cm]{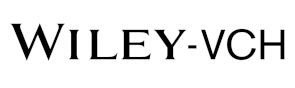}}

\title{Transient Chirp Dynamics in Terahertz Quantum Cascade Lasers}

\maketitle

% Author: Please give full first and last names for authors and include * after the name of all corresponding authors

\author{Xianglong Bi}
\author{Xuhong Ma}
\author{Wenjian Wan}
\author{Binbin Liu}
\author{Guibin Liu}
\author{Ziping Li}
\author{Yanming Lu}
\author{Zhiwei Qin}
\author{Yunxiang Zhu}
\author{Ziyu Guo}
\author{J. C. Cao}
\author{Hua Li}

% Dedication

\dedication{}

% Affiliations: Please provide adacemic titles (Prof. or Dr.) for all authors where applicable, and include an institutional email address for all corresponding authors
\begin{affiliations}
X. Bi, Dr. X. Ma, Prof. W. Wan, Dr. B. Liu, G. Liu, Prof. Z. Li, Y. Lu, Z. Qin, Y. Zhu, Z. Guo, Prof. J. C. Cao, Prof. H. Li\\
National Key Laboratory of Materials for Integrated Circuits and Key Laboratory of Terahertz Solid State Technology, Shanghai Institute of Microsystem and Information Technology\\ Chinese Academy of Sciences \\
865 Changning Road, Shanghai 200050, China\\
Email: maxuhong@mail.sim.ac.cn, jccao@mail.sim.ac.cn, hua.li@mail.sim.ac.cn

X. Bi, G. Liu, Y. Lu, Z. Qin, Y. Zhu, Prof. J. C. Cao, Prof. H. Li\\
Center of Materials Science and Optoelectronics Engineering \\
University of Chinese Academy of Sciences\\
Beijing 100049, China

Dr. X. Ma, Dr. B. Liu\\
Chongqing Key Laboratory of Precision Optics\\
Chongqing Institute of East China Normal University\\
Chongqing 401120, China
\end{affiliations}

% Keywords: Please provide a minimum of three and a maximum of seven keywords, separated by commas


% Abstract should be written in the present tense and impersonal style (i.e., avoid we), and be at most 200 words long
\begin{abstract}
Laser frequency chirp is a ubiquitous dynamical process in semiconductor lasers, vital for frequency-modulated photonic systems. In the mid-infrared (MIR) and terahertz (THz) ranges, quantum cascade lasers (QCLs) are ideal sources with high power, narrow linewidth and compact size. While chirp dynamics in MIR QCLs have been studied, the transient chirp behavior of THz QCLs---particularly the thermal chirp on microsecond to millisecond timescales---remains largely unexplored. Here, we experimentally investigate transient thermal chirp dynamics in single-mode THz QCLs via an on-chip heterodyne scheme. Twin monolithically integrated single-mode QCLs are used: one pulsed QCL as the device under test, and one continuous-wave (CW) QCL serving as both local oscillator (LO) and ultrafast THz detector. The frequency chirp is mapped to the radio-frequency (RF) domain by heterodyne down-conversion. By varying current and temperature, we observe three distinct chirp features: unidirectional down-chirp, V-shaped chirp, and unidirectional up-chirp. A two-node thermal model reproduces the dynamics with good agreement with experiments. Chirp dynamics in the multi-mode regime are also identified, showing the potential for sensitive dynamic spectral characterization. These findings deepen the understanding of THz QCL thermal chirp mechanisms and support applications in THz frequency combs, frequency-modulated continuous-wave (FMCW) radar, and high-speed coherent communications.
\end{abstract}

% Text: Please use section headings and subheadings as specified below. For communications, all section headings apart from Experimental Section should be removed
% Please make the first reference to a display item bold: \textbf{Figure 1}
% Do not abbreviate Figure, Equation, etc.; display items are always singular, i.e., Figure 1 and 2.
% Equations are always singular, i.e., Equation 1 and 2, and should be inserted using the {equation} environment, not as graphics
% Please do not use footnotes in the text, additional information can be added to the Reference list.

\section{Introduction}
Frequency chirp, defined as the time-dependent fluctuation of the optical carrier frequency, is inherent to nearly all laser systems and exerts a crucial influence on their performance. On the one hand, the presence of unregulated frequency chirp in certain applications can lead to severe degradation of system accuracy and reliability, including increased bit error rates in coherent optical communication\cite{agrawal1986effect, goi201211}, blurred fingerprint peaks in molecular detection\cite{shipp2017raman, markmann2023frequency}, and reduced spatial resolution in non-destructive imaging\cite{arora2014pulse, su2021high}. On the other hand, well-regulated frequency chirp serves as a valuable tool in numerous laser-based applications, such as enabling wideband frequency modulation in frequency-modulated continuous wave (FMCW) radar\cite{benson2016digital, 2020Massively, lukashchuk2022dual} and facilitating pulse compression in chirped pulse amplification (CPA) for ultra-short laser pulses\cite{strickland1985compression, han2020generation}. Given the critical impact of chirp on both system performance and functional applications, a comprehensive understanding of chirp dynamics is essential for emerging laser systems.

Notably, semiconductor-based quantum cascade lasers (QCLs)\cite{faist1994quantum, kohler2002terahertz, yao2012mid, 2026294} have emerged as versatile emitters across the mid-infrared (MIR) to terahertz (THz) spectral regions, thanks to their high output power, broad frequency tunability and compact footprint. Distinct from conventional semiconductor lasers relying on interband electron-hole recombination, QCLs achieve laser emission through the resonant intersubband transition of electrons in the artificially designed quantum well superlattice structure\cite{gmachl2001recent, capasso2010high}, which enables flexible tailoring of the emission wavelength across the MIR to THz range via band structure engineering. As electrically pumped semiconductor devices, QCLs have garnered extensive research attention over the past decades.
In addition to the extensive research on steady-state frequency performance\cite{qin2009tuning, curwen2019broadband, kundu2020wideband} and and its enhancement\cite{okajima2003experimental, li2019graphene, zhao2021active, li2023terahertz, Li:18, zhao2024stabilized, liu2025farey},the transient chirp dynamics of QCLs have emerged as a meaningful and increasingly attractive topic for understanding the full frequency behavior of these versatile sources. In addition to the extensive research on steady-state frequency performance and its enhancement, the transient chirp dynamics of QCLs have emerged as a meaningful and increasingly attractive topic for understanding the full frequency behavior of these versatile sources. In the single-mode QCL aspect, Kundu et al. investigated the ultrafast switching process in THz coupled-cavity QCLs using an injection seeding technique and explained its underlying physics through a full multi-mode carrier and photon transport model\cite{kundu2018ultrafast}. Qi et al. reported the transient instabilities in a THz single-mode QCL under optical feedback and performed thorough simulations based on a reduced rate equation model\cite{qi2021observation}. For QCL frequency combs, the intrinsic frequency modulation (FM) represents another important class of chirp-related dynamics. Experimentally, this chirp behavior can be measured using the well-known shifted wave interference Fourier transform spectroscopy (SWIFTS) technique\cite{burghoff2014terahertz, burghoff2015evaluating}. Theoretically, Maxwell--Bloch equations provide a powerful framework for modeling and simulating QCL comb systems. Extensive investigations have been carried out on the chirp dynamics in QCL combs, covering aspects such as cavity engineering\cite{seitner2025aspects}, passive mode-locking\cite{burghoff2020unraveling}, self-FM comb dynamics\cite{roy2024self}, and time-domain comb generation\cite{tzenov2016time}.

These state-of-the-art measurement tools and carrier--photon frameworks have provided satisfactory interpretations of QCL transient dynamics within the ps--ns range. However, thermal accumulation is negligible on these short timescales, rendering such approaches inadequate for describing frequency dynamics on $\mu$s--ms timescales where thermal effects become prominent. Several investigations have addressed thermal chirp behavior in the MIR region. For instance, Michael et al. first characterized the frequency chirp of MIR distributed feedback (DFB) QCLs using high-resolution Fourier-transform spectrometers\cite{mcculloch2003highly}; subsequent studies further probed frequency chirp in pulsed MIR QCLs and validated their applications in laser absorption spectroscopy (LAS)\cite{2010Molecular, nair2022extended, lin2023real} and laser cavity temperature measurements\cite{gundogdu2018time}. By contrast, the thermal chirp dynamics of THz QCLs during the critical startup phase remain largely unexplored. This unresolved research gap hinders the full characterization of THz QCL dynamic frequency properties and limits their utilization in practical terahertz applications. This gap persists primarily because of the lack of high-performance coherent detectors in the THz band\cite{rogalski2003infrared, sizov2010thz, dhillon20172017}. Existing THz coherent detectors have inherent limitations: Schottky barrier diode (SBD) mixers\cite{danylov2015phase, yang2016terahertz} lack sufficient sensitivity, while superconducting hot-electron bolometers (HEBs)\cite{richter2008terahertz, yang2016terahertz} and quantum well infrared photodetectors (QWIPs)\cite{rogalski2003quantum, li20176} require cryogenic cooling, making systems bulky. Notably, THz QCLs themselves can act as high-sensitivity detectors with excellent spectral matching and on-chip integration compatibility\cite{liu2025terahertz, bi2025terahertz}, offering a feasible solution to this bottleneck and facilitating compact THz detection systems for studying their transient chirp dynamics.

In this work, we experimentally probe the thermal chirp dynamics of a pulsed single-mode THz QCL by leveraging an on-chip heterodyne detection scheme. Specifically, twin single-mode THz QCLs are monolithically integrated on the same chip, where one is driven in pulsed mode as the signal source to exhibit chirp dynamics, and the other operates under continuous-wave (CW) conditions, simultaneously serving as a local oscillator (LO) and an ultrafast detector for coherent signal acquisition. By systematically varying the operating current and temperature of the pulsed THz QCL, we characterize the evolution of its transient chirp dynamics and identify three characteristic features in the down-converted RF beat signals. To gain fundamental physical insights into the observed chirp behaviors, a two-node thermal model is employed to simulate the THz QCL chirp dynamics under a typical operating condition, with the simulation results showing good consistency with experimental observations. Furthermore, we identify chirp dynamics in the multi-mode regime, which demonstrates the potential of our on-chip heterodyne detection approach for highly sensitive dynamic spectral characterization of THz QCLs. This work not only advances the fundamental understanding of thermal chirp behaviors in THz QCLs but also establishes an on-chip heterodyne framework that can be extended to future studies of transient frequency dynamics in both single-mode and multi-mode regimes.

\section{Results}
\label{sect:Materials}
\begin{figure}[htbp]
\centering
\includegraphics[width=0.98\linewidth]{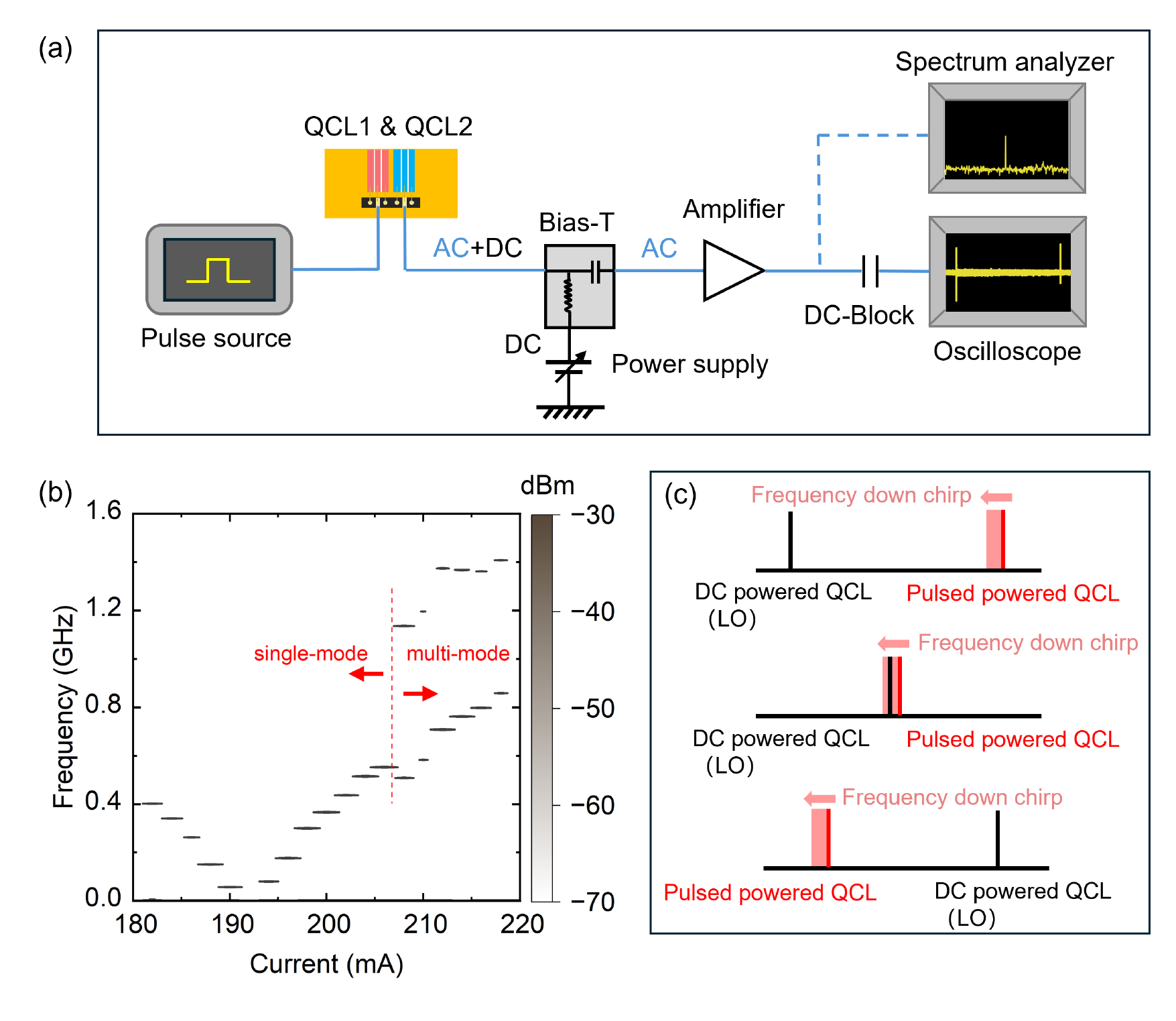} 
\caption 
{(a) Experimental setup for heterodyne detection of chirp dynamics in THz QCLs. The twin QCLs (QCL1 and QCL2) are monolithically fabricated on a single substrate to form an on-chip integrated dual-device. QCL1 is driven by a pulse source to induce frequency chirp, while QCL2 is CW-driven and acts as a local oscillator (LO) for heterodyne detection. A DC-block is used to block leaked DC current and only transmit the beat-frequency chirp signal. (b) Steady-state beat-frequency mapping of QCL1 measured via a spectrum analyzer with a resolution bandwidth (RBW) of 100 kHz and a video bandwidth (VBW) of 50 kHz. (c) Schematic illustration of the relative lasing frequency positions of the twin QCLs. The lasing frequency of the CW-driven QCL is fixed and serves as the LO, whereas the lasing frequency of the pulse-driven QCL exhibits an ``intrinsic downward frequency chirp". From top to bottom, the schematics represent three scenarios of the relative lasing frequency relationship between the two QCLs, which we term the ``unidirectional down-chirp", ``V-shaped chirp", and ``unidirectional up-chirp" scenarios, respectively. These scenarios correspond to the three distinct beat-frequency chirp signals in the RF band observed in the experiment.} 
\label{f1}
\end{figure}

Figure \ref{f1}(a) illustrates the experimental setup for heterodyne detection of chirp dynamics in THz QCLs.
The twin THz QCLs employed in this work were both designed with a 1 mm cavity length and a 150 µm ridge width, and monolithically integrated on a single substrate with a separation of 1.6 mm, forming an on-chip dual-emitter system. The coupling between the two lasers is mediated via the shared GaAs substrate, as the single-plasmon waveguide geometry allows a significant portion of the optical mode to extend into the substrate.\cite{li2019chip}
 Thanks to their short cavity design, both devices operated reliably in a single-mode regime under our experimental conditions, with their lasing frequencies matched to enable efficient heterodyne beat-frequency generation. Their emission spectra are presented in Fig. S1 of the Supplementary Material, as characterized via Fourier transform infrared (FTIR) spectroscopy. For the experiment, QCL1 was driven by a pulse source to induce frequency chirp, while QCL2 was biased with a CW power supply to generate a LO frequency. Benefiting from the intrinsic ultrafast carrier relaxation dynamics of THz QCLs, QCL2 acts as a high-speed mixer to heterodyne-detect and down-convert the chirp dynamics of QCL1 to the microwave band\cite{li2015dynamics, li2019toward, ma2025self}. The down-converted microwave signal was first extracted via the AC port of a bias-T, amplified by 30 dB using a microwave amplifier, passed through a DC-block, and finally acquired by a high-speed real-time oscilloscope. An additional measurement branch connected to a spectrum analyzer was used to capture the beat-frequency signal of the two QCLs under steady-state conditions, where both devices were biased for CW operation. A key distinction in the measurement setup is that the spectrum analyzer is dedicated to detecting the steady-state beat-frequency signal with both QCLs operating under CW drive, whereas the oscilloscope captures the down-converted chirp signal in the configuration where QCL1 is pulse-driven and QCL2 remains CW-driven.

Fig. \ref{f1}(b) presents the beat-frequency mapping of QCL1 measured via the spectrum analyzer, which depicts the current-tuning characteristics of QCL1 under CW operation with its driving current increased stepwise from 180 mA to 220 mA, while QCL2 was maintained at a fixed CW driving current of 230 mA. The heat sink temperature was stabilized at 6.5 K. In our previous work, we identified that the lasing frequency of THz QCL blueshifts with increasing driving current\cite{guan2021frequency}. Accordingly, the mapping illustrates the evolution of QCL1’s lasing frequency: as the current increases, the frequency first rises to approach that of QCL2, then gradually deviates with further current increment. When the current increases from 180 mA to 208 mA, QCL1’s lasing frequency increases linearly with current, exhibiting a tuning rate of 38.55 MHz/mA. This linear frequency-current relationship is evidenced by the single beat-frequency signal and its linear trend with current in the mapping diagram. At a driving current of 192 mA, the beat frequency reaches zero, indicating that the lasing frequencies of the two QCLs are approximately coincident at this bias point. As the current is further increased above 208 mA, QCL1 exhibits two distinct lasing frequencies with an approximate spacing of 600 MHz, accompanied by slight redshift jumps. This observation directly demonstrates the transition of QCL1 from a single-mode to a multi-mode operating regime, where the linear correlation between the device’s lasing frequency and driving current no longer holds. The frequency tuning behavior is in good agreement with the emission spectra results presented in Fig. S1 of the Supplementary Material. Correspondingly, Fig. S2 of the Supplementary Material depicts the current tuning characteristics of QCL2.
 
Depending on the relative positional differences between the lasing frequencies of the two QCLs, three expected chirp dynamic heterodyne detection processes are illustrated from top to bottom in Fig. \ref{f1}(c), which we term the unidirectional down-chirp, V-shaped chirp, and unidirectional up-chirp scenarios, respectively. The top plot depicts the unidirectional down-chirp scenario, where the frequency chirp process of the pulse-driven QCL lies entirely above the lasing frequency of the LO QCL. As the pulse-driven QCL undergoes its intrinsic downward frequency chirp, the chirped frequency continuously approaches the LO frequency, corresponding to a unidirectional down-chirped signal in the RF band. The middle plot illustrates the V-shaped chirp scenario, where the downward frequency chirp process of the pulse-driven QCL crosses the lasing frequency of the LO QCL from start to finish. The corresponding down-converted RF chirp signal undergoes a transition from downward chirping, through zero frequency, to reversed upward chirping, forming a characteristic V-shaped profile. The bottom plot represents the unidirectional up-chirp scenario, where the downward frequency chirp process of the pulse-driven QCL occurs entirely at frequencies lower than the LO QCL’s lasing frequency. Here, the intrinsic downward chirp of the THz QCL causes its frequency to move further away from the LO frequency, resulting in a unidirectional upward chirping profile in the down-converted RF signal. It is critical to note that this classification is based solely on the relative frequency position of the pulse-driven QCL with respect to the LO QCL, rather than the intrinsic chirp direction of the THz QCL itself—intrinsically, the THz QCL exhibits a pure downward frequency chirp in all cases.

\begin{figure}[t]
\centering
\includegraphics[width=0.98\linewidth]{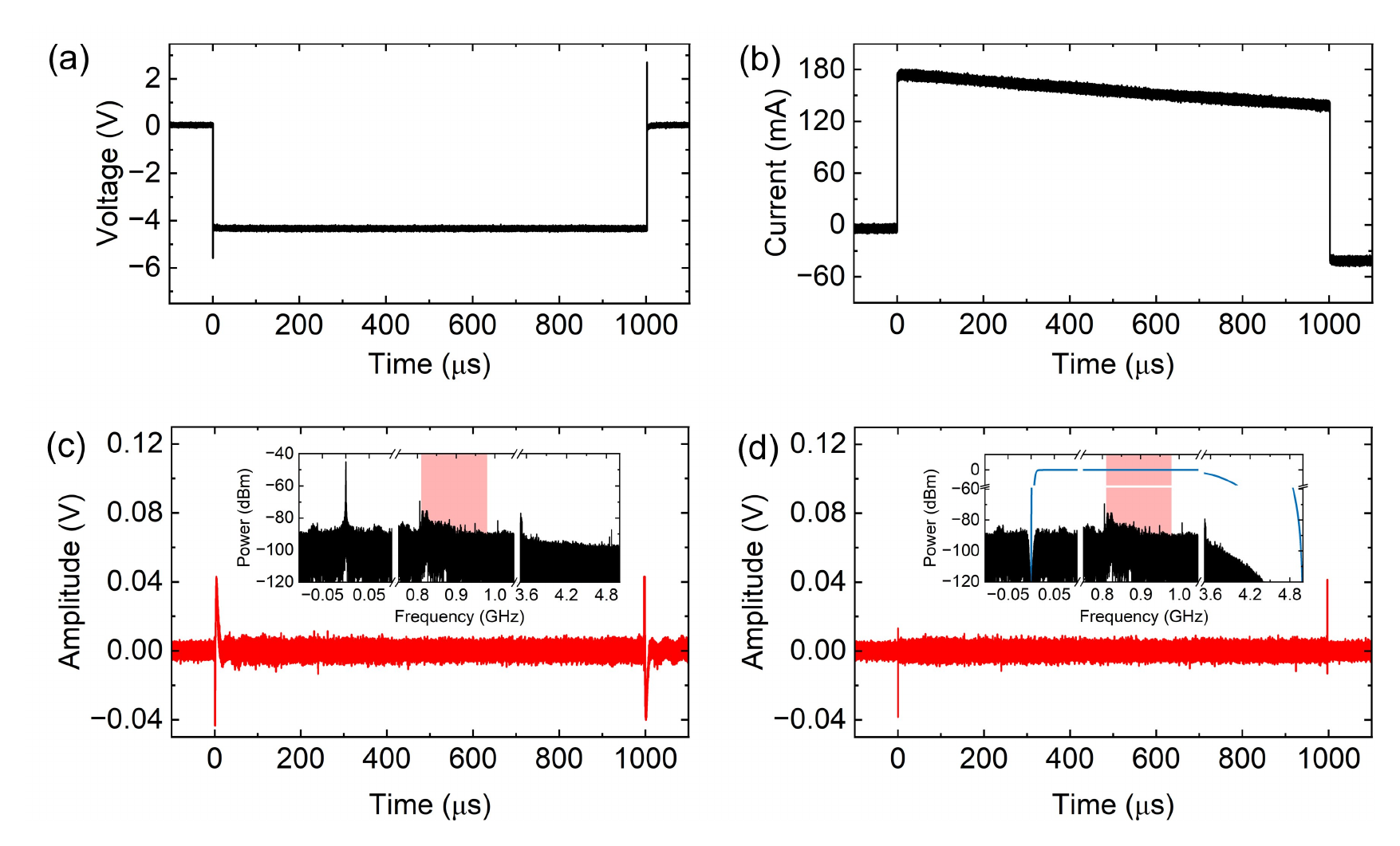} 
\caption 
{(a) Pulsed voltage and (b) pulsed current applied to QCL1. (c) Time trace of the down-converted beat-frequency chirp signal in the RF band acquired by the oscilloscope and (d) that after band-pass filtering. The insets in (c) and (d) show the fast Fourier transform (FFT) results of the corresponding time traces, respectively. All measurements were performed at a fixed heat sink temperature of 11.3 K with CW-driven QCL2 operated at 226 mA.} 
\label{f2}
\end{figure}

In our experiment, the pulse source (Keysight 8114A) is used to drive QCL1 and induce its chirp dynamics. The QCL is continuously driven by pulses with a duration of 1000 $\mu$s and a duty cycle of 10\%, where the voltage and current signals of the QCL during the pulse duration are displayed in Figs. \ref{f2}(a) and \ref{f2}(b), respectively. Owing to the non-ideal characteristics of the pulse source, the voltage and current signals deviate from ideal square waveforms. Specifically, the voltage signal exhibits overshoots at the rising and falling edges due to LC oscillations induced by parasitic parameters in the pulse-driven circuit, while the current signal shows a ramp-down trend from 180 mA to 140 mA during the pulse duration because of the gradual attenuation of the power supply’s output capability under constant-voltage operation\cite{hanigovszki2004novel, chen2013soft}. For the sake of simplicity, in subsequent discussions, the pulse-driven current applied to the QCL is denoted by the value corresponding to the midpoint of the pulse duration (160 mA at 500 $\mu$s in this case). Note that the current in Fig. \ref{f2}(b) exhibits a negative value after the falling edge. This is a transient artifact of the current transformer at the falling edge, caused by the inherently AC-coupled nature of the Pearson current monitor (Model 6027) we use. Physically, after the falling edge, the current has already returned to zero; the negative signal does not represent a real reversal of the device current.

Fig. \ref{f2}(c) shows the temporal waveform of the down-converted beat-frequency chirp signal in the RF band acquired by the oscilloscope. At the arrival of the pulse edges, the time-domain signal exhibits a transient abrupt change followed by a certain degree of oscillation. The inset plot corresponds to the FFT result of the time-domain signal, where the pink region in the middle reflects the chirped region of the down-converted signal. Across the frequency band spanning 810 MHz to 980 MHz, the chirped signal displays a gradually decreasing power profile. The strong peak at 0 Hz on the left indicates that significant DC leakage still exists, even though a DC-block was incorporated into the measurement link. To further filter out various interference signals, a digital band-pass filter (passband range: 5 kHz–3.5 GHz) was applied to process the time-domain signal. As shown in Fig. \ref{f2}(d), the transient fluctuations and oscillations of the time-domain signal at the pulse edges are effectively suppressed after filtering. The FFT result in the inset also demonstrates the filtering effect of the band-pass filter on both DC components and high-frequency components above 3.5 GHz. The inset additionally presents the amplitude-frequency characteristic curve of the band-pass filter, represented by a blue line.

\begin{figure}[htbp]
\centering
\includegraphics[width=0.98\linewidth]{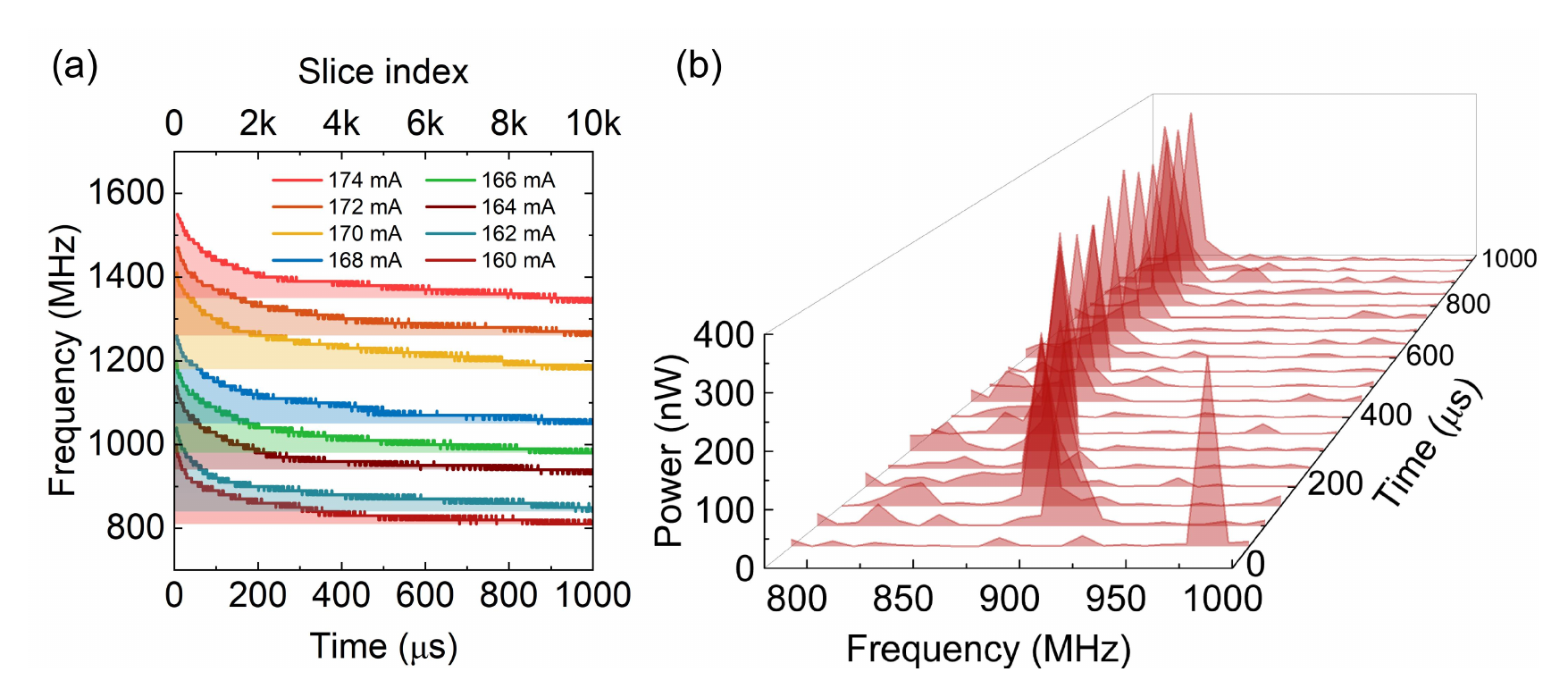} 
\caption 
{(a) Frequency chirp dynamics of QCL1 under different pulse-driven current conditions. (b) Combined spectrograms of QCL1 at 160 mA pulse-driven current (selected at 50 $\mu$s intervals from 10000 total temporal slices). All measurements were performed at a fixed heat sink temperature of 11.3 K with CW-driven QCL2 operated at 226 mA.} 
\label{f3}
\end{figure}

Fig. \ref{f3}(a) further presents the frequency chirp dynamics of QCL1 under various pulse-driven current conditions, where the pulse-driven current of QCL1 is increased from 160 mA to 174 mA in 2 mA steps. Following acquisition by an oscilloscope and filtering with the digital band-pass filter (introduced in Fig. \ref{f2}), the temporal waveforms were analyzed via the short-time Fourier transform (STFT). Specifically, the 1000 $\mu$s pulse duration was divided into 10,000 temporal slices with a 0.1 $\mu$s duration per slice. By identifying the frequency peaks of these slices, the frequency chirp dynamics were extracted. A detailed elaboration of the underlying principle for peak identification in temporal slices is provided in the Methods section. Taking the 160 mA curve as an example, the overall chirp range spans approximately 810-1000 MHz over the 1000 $\mu$s pulse duration, exhibiting a continuous downward chirp (frequency decreasing) throughout this period. The frequency drops relatively rapidly in the early stage, while the chirp rate gradually slows down as time elapses, showing an overall ``fast-then-slow" downward frequency trend. This can be attributed to the fact that with the continuation of current driving, the internal thermal field of the active region gradually transitions from a highly unstable state to a stable one, and the refractive index of the active region stabilizes accordingly. This chirping behavior corresponds to the power distribution of the down-converted chirp signal indicated by the pink region in Fig. \ref{f2}(c). As the pulse-driven current increases, the curves shift entirely toward the higher-frequency regime, forming a family of curves with similar variation patterns. Considering the curve frequencies at the 500 $\mu$s time point, as the pulse-driven current increases from 160 mA to 174 mA, the frequency rises from 830 MHz to 1380 MHz, corresponding to a current tuning rate of approximately 39.29 MHz/mA, which is comparable to the DC current tuning rate presented in Fig. \ref{f1}(b). The combination of current tuning and frequency chirp enables gapless frequency coverage ranging from 810 MHz to 1550 MHz in total. This seamless spectral span holds considerable practical potential for frequency-modulated (FM) communication and high-resolution molecular detection applications in the THz regime. In Fig. \ref{f3}(b), spectral slices of QCL1 were selected at 50 $\mu$s intervals from the 10,000 total temporal slices and combined with their frequency axes aligned. The peak power of these spectral slices is approximately 7 nW, with a frequency resolution of 10 MHz determined by the 0.1 $\mu$s temporal slice duration. For a more extensive set of temporal slices characterizing the frequency chirp dynamics, refer to Fig. S3 in the Supplementary Material.

\begin{figure}[htbp]
\centering
\includegraphics[width=0.98\linewidth]{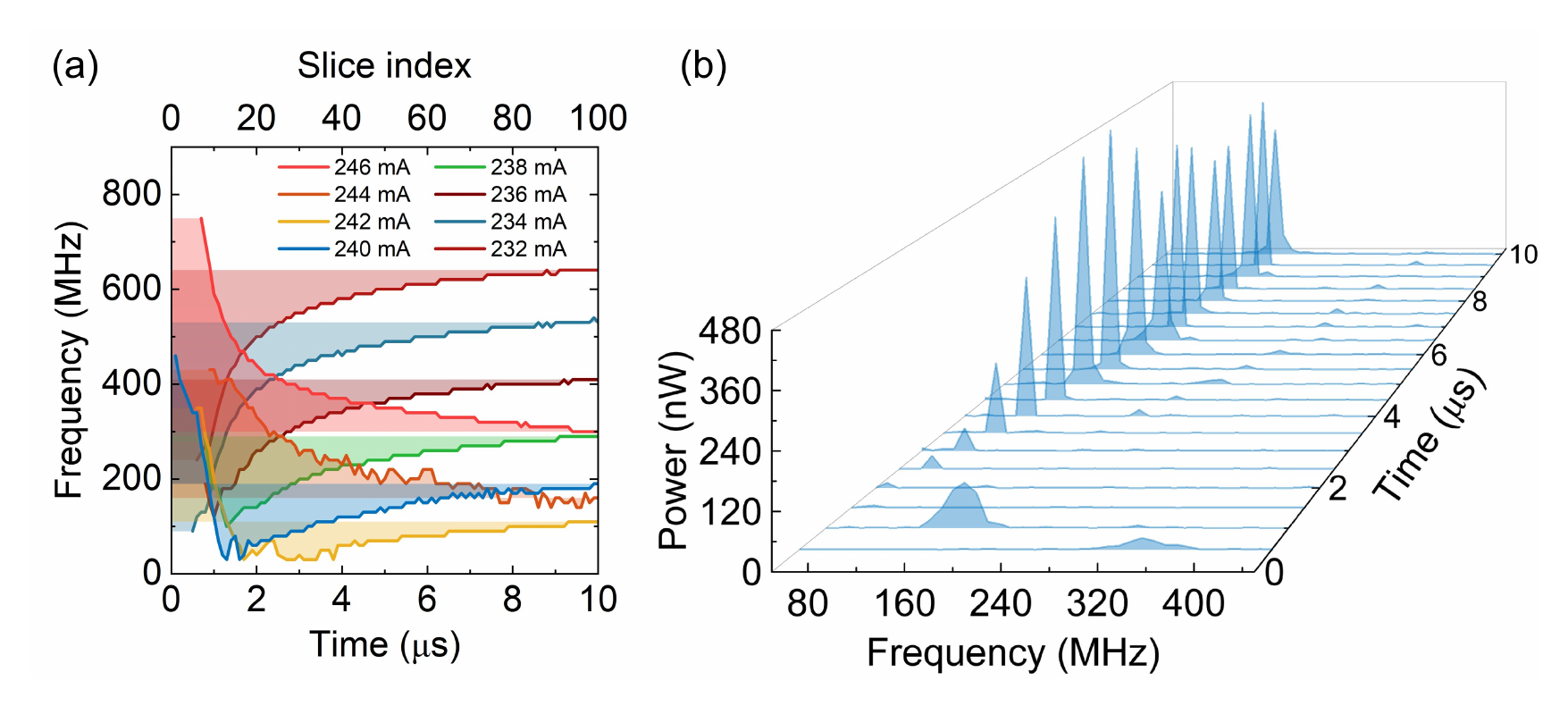} 
\caption 
{(a) Frequency chirp dynamics of QCL2 under different pulse-driven current conditions. (b) Combined spectrograms of QCL2 at 240 mA pulse-driven current (selected at 0.5 $\mu$s intervals from 10000 total temporal slices). All measurements were performed at a fixed heat sink temperature of 11.3 K, with CW-driven QCL1 operated at 195 mA.} 
\label{f4}
\end{figure}

 Our on-chip dual single-mode device features a symmetric architecture, whereby the frequency chirp dynamics of QCL2 can also be characterized under pulse-driven operation using the identical experimental setup. To further validate the robustness of our experimental results, we present the frequency chirp dynamics of QCL2 in Fig. \ref{f4}, with the key difference from the QCL1 measurements lying in the 10 $\mu$s pulse duration (10\% duty cycle) employed for QCL2. Under this operating condition, QCL2 exhibits an identical continuous downward chirp trend to QCL1 but with a considerably higher chirp rate. As shown in Fig. \ref{f4}(a), when biased with a 246 mA pulse-driven current, QCL2 exhibits a total frequency chirp range of 450 MHz, shifting from 750 MHz to 300 MHz. Following the application of the current pulse, QCL2 requires approximately 0.7 
$\mu$s to achieve stable lasing, which defines the origin of the chirp curve. A comprehensive set of temporal slices characterizing the full frequency chirp dynamics is provided in Fig. S4 of the Supplementary Material, which details the evolution of lasing frequency generation subsequent to pulse excitation. When we adjust the pulse current, QCL2’s down-chirp curve undergoes the three distinct processes predicted in Fig. \ref{f1}(c). For instance, under the 240 mA pulsed current condition, the curve first down-chirps to zero frequency and then reverses to an up-chirp, indicating that QCL2’s down-chirped frequency crosses QCL1’s lasing frequency during this interval, which corresponds to the V-shaped chirp scenario described in Fig. \ref{f1}(c).

 Fig. \ref{f4}(b) presents the combined spectral slices selected at 0.5 $\mu$s intervals. It can be seen that the first slice, corresponding to the 0.5 $\mu$s timestamp, spans approximately 80 MHz, which indicates that the frequency chirp range reaches 80 MHz within the 0.1 $\mu$s slice duration. Owing to energy distribution across this broad frequency band, the peak power of the entire spectral slice remains relatively low. When time reaches 1 $\mu$s, the chirp rate decreases, and the corresponding spectral slice exhibits a narrower coverage range and higher peak power. As the chirp frequency approaches zero (corresponding to 1.5-2 $\mu$s), the spectral slices show weak peak powers. This phenomenon is attributed to the bandwidth limitations of the DC-Block and band-pass filter integrated into the measurement setup. Owing to the power suppression applied to frequencies below 5 kHz, it is difficult to accurately locate the peak of the chirp signal near zero frequency. Consequently, in Fig. \ref{f4}(a), the four chirp curves (236-242 mA) that undergo a transition at the zero point exhibit ambiguity in the vicinity of zero frequency. As time further exceeds  2 $\mu$s, the chirp frequency moves away from zero frequency, as shown in Fig. \ref{f4}(b); meanwhile, the signal power gradually increases, and the power distribution of these slices narrows to the range of 20–40 MHz. This chirping behavior of the beat-frequency down-converted signal is described in greater detail in Fig. S5 of the Supplementary Material. The power differences across Figs. S3–S5 stem from the intensified beat frequency effect induced when the terahertz chirped pulse frequency approaches the LO frequency\cite{barkan2004linewidth, ren2011high}. These two effects (the enhanced beat frequency signal near zero frequency and the suppression around zero frequency) act in combination, resulting in the beat-frequency chirp signal in Fig. \ref{f4}(b) achieving a signal intensity of hundreds of nWs within the 100–200 MHz range.
 
\begin{figure}[htbp]
\centering
\includegraphics[width=0.98\linewidth]{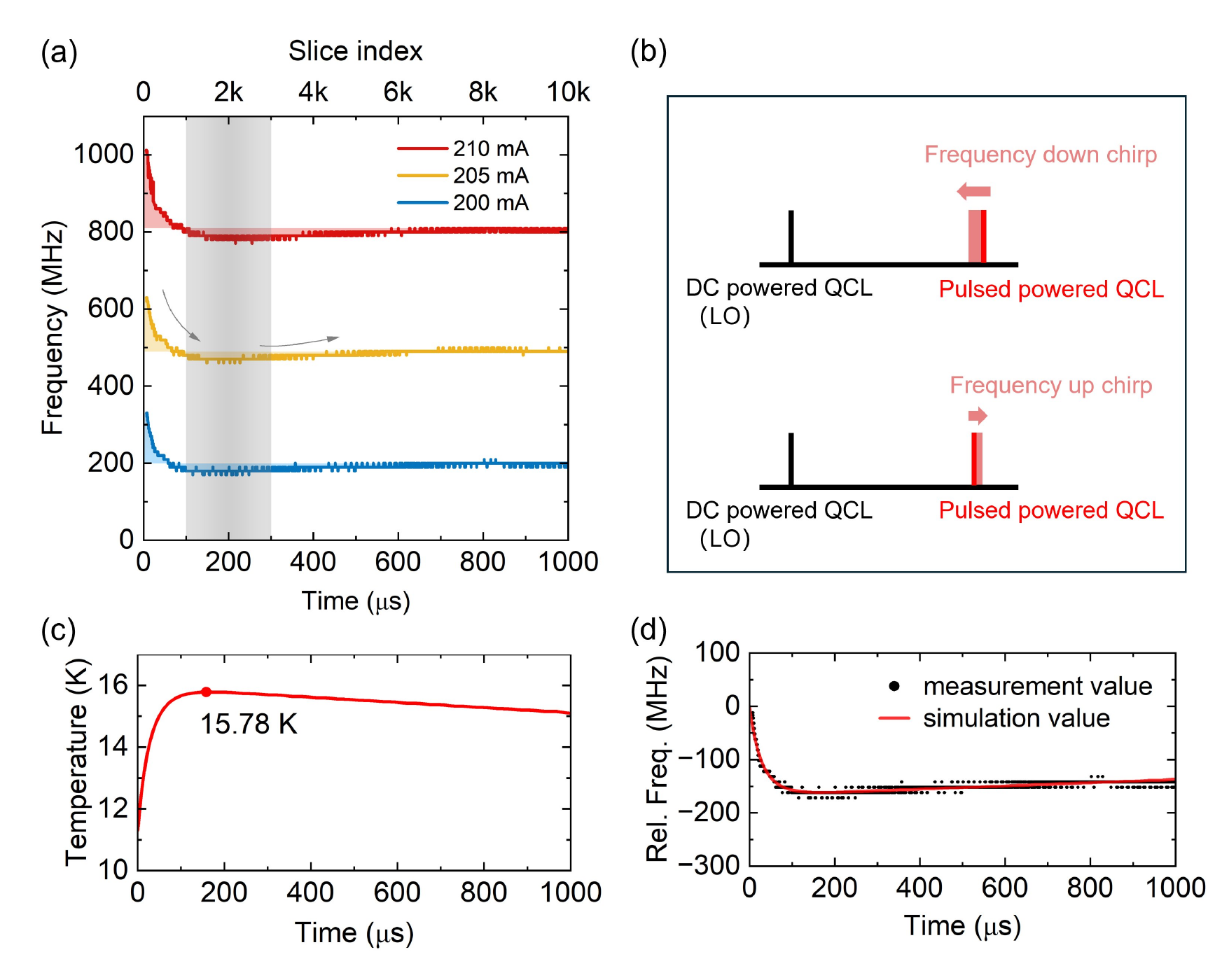} 
\caption 
{(a) Experimental results of non-unidirectional chirp dynamics for QCL2, measured at a fixed heat sink temperature of 11.3 K with CW-driven QCL1 biased at 175 mA. (b) Schematic illustration of non-unidirectional chirp in a pulse-driven QCL. (c) Simulated temporal evolution of the active region temperature in the QCL under a 200 mA pulse-driven current. (d) Comparison of simulated QCL frequency chirp dynamics (red solid line) with experimental measured values (black dots) under a 200 mA pulse-driven current.} 
\label{f5}
\end{figure}

Later, for a more comprehensive investigation of QCL2’s chirp dynamics under varying operating conditions, we adjusted the current and temperature parameters and remeasured QCL2’s chirp curve under 1000 $\mu$s pulsed excitation. Fig. \ref{f5}(a) demonstrates that the chirp curves deviate from a purely unidirectional trend under the three pulsed current conditions (200, 205, and 210 mA): after down-chirping to a minimum frequency, a slight upward chirp emerges. The gray shaded region (100–300 $\mu$s) indicates the interval where the chirp direction reversal occurs; the ambiguity in the exact reversal timestamp arises from both the slow chirp rate within this period and the 10 MHz frequency resolution limitation of the measurement system. What's more, the chirp rate and chirp range under these conditions are substantially smaller than those in Fig. \ref{f4}(a). This phenomenon can be attributed to the fact that a lower-amplitude pulsed current induces a smaller temperature jump in the QCL active region, which in turn results in a relatively weaker frequency chirp\cite{fischer2014intermittent, nair2022extended}. Fig. \ref{f5}(b) depicts the schematic of non-unidirectional chirp behavior. The pulse-driven QCL exhibits two distinct chirp trends: frequency down-chirp (upper panel) and frequency up-chirp (lower panel). Specifically, the emission frequency of the pulsed QCL deviates from that of the CW-driven LO in both downward and upward directions—this bidirectional variation directly reflects the non-unidirectional chirp characteristic observed in our experiments. 

To further validate the experimentally observed non-unidirectional chirp behavior and understand its trend, we performed a simulation on the chirp frequency dynamics of the Fabry–Pérot (FP) laser. The chirp frequency of the FP laser and the temperature of its active region satisfy the following relationship\cite{gundogdu2018time, geng2018effects}:  
\begin{equation}
\Delta \nu = -\nu_{0} \alpha_{\mathrm{eff}} \Delta T = -\nu_{0} \alpha_{\mathrm{eff}} \left[ T(t) - T_{0} \right]
\end{equation}
where $\Delta \nu$ denotes the chirp frequency, $\nu_0$ is the absolute emission frequency of the laser, $\alpha_{\mathrm{eff}}$ is the effective thermo-optic coefficient, $T(t)$ is the real-time temperature of active region, and $T_0$ is the initial temperature of the active region.
To solve for the temperature variation in the active region, we established a two-node thermal model governed by the following heat transfer equations:
\begin{align}
C_1 \frac{dT_1}{dt} &= P_{\text{heat}}(t) - \frac{T_1 - T_2}{R_{12}}\\
C_2 \frac{dT_2}{dt} &= \frac{T_1 - T_2}{R_{12}} - \frac{T_2 - T_{\text{sink}}}{R_{\mathrm{2s}}}
\end{align}
Here, the two nodes, labeled 1 and 2, correspond to the active region and substrate packaging, respectively. $T_1$ and $T_2$ denote their respective temperatures, while $C_1$ and $C_2$ represent their heat capacities. $T_{\text{sink}}$ refers to the temperature of the heat sink, $P_{\text{heat}}(t)$ is the heating power induced by pulsed current excitation, $R_{12}$ and $R_{\mathrm{2s}}$ represent the thermal resistance corresponding to the node 1–node 2 interface and the node 2–heat sink interface, respectively. 

Using the model described above, we performed simulations under 200 mA pulsed current excitation. All parameters employed in the simulations are listed in Table S1 of the Supplementary Material. The simulated temporal evolution of the active region temperature is presented in Fig. \ref{f5}(c). It can be observed that the temperature of the active region exhibits an overall trend of rapid initial rise followed by gradual decrease. The initial temperature is set at 11.3 K, consistent with the heat sink temperature. At 156 $\mu\text{s}$, the temperature peaks at 24.79 K and then gradually decreases to 22.7 K at 1000 $\mu\text{s}$. This thermal behavior can be attributed to the following mechanisms: As soon as the rising edge of the pulse arrives, the active region undergoes rapid ohmic heating. With the increase in the temperature of the active region and the ramp-down of the current, the temperature rise rate of the active region slows down gradually. At 156 $\mu\text{s}$, thermal equilibrium between heating and heat dissipation is reached. Subsequently, the current continues to decrease in a ramp profile; as heat dissipation exceeds heating, the temperature of the active region decreases slowly. The corresponding frequency chirp is plotted as the red curve in Fig. \ref{f5}(d), which shows excellent agreement with the experimental chirp measurements represented by the black dots. Overall, the thermal dynamics revealed by the simulation provide a clear physical interpretation for the observed frequency chirp behavior in the experiments.

\begin{figure}[t]
\centering
\includegraphics[width=0.98\linewidth]{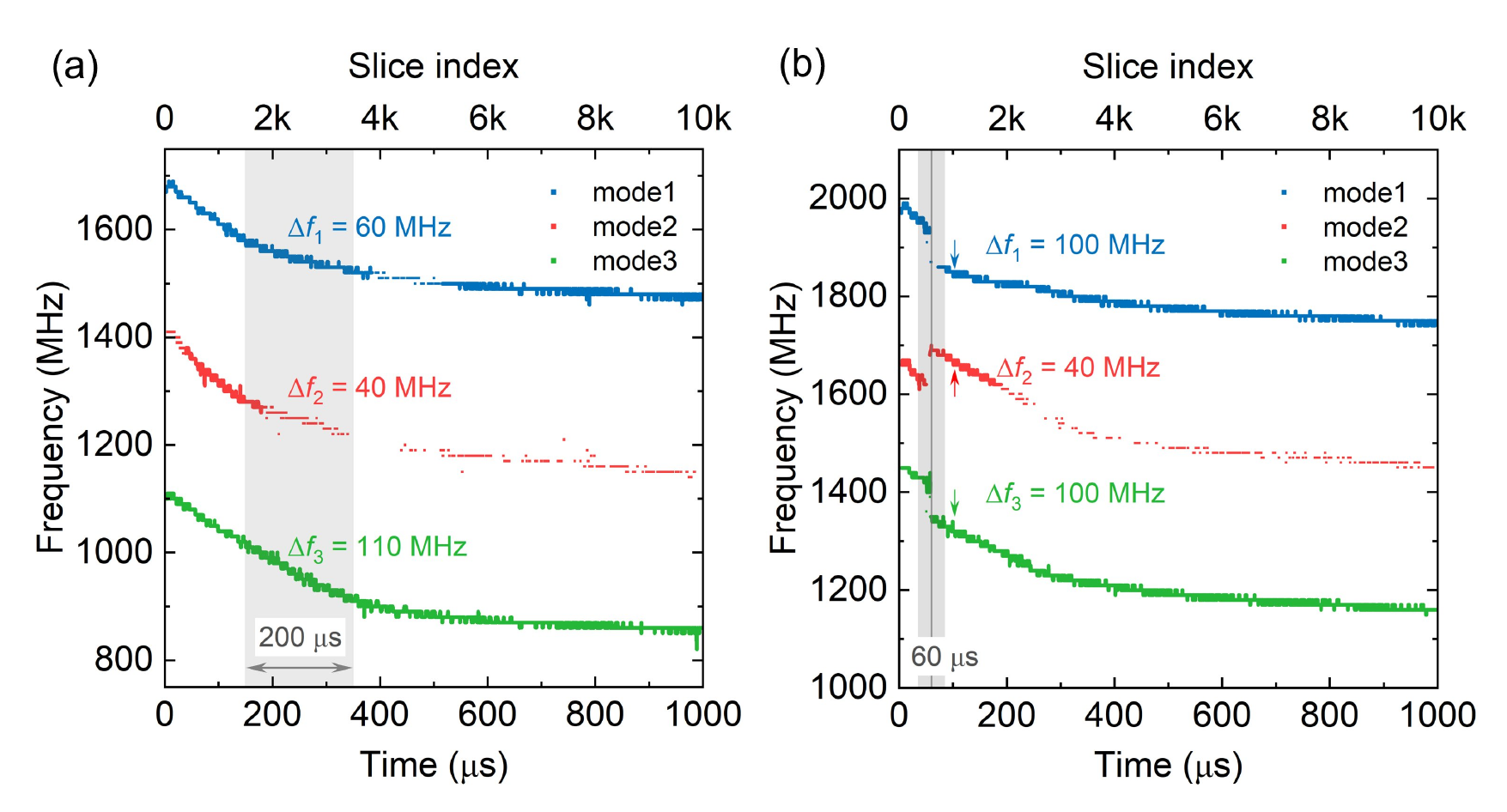} 
\caption 
{(a) Continuous variation of chirp frequency for QCL1 under a 180 mA pulse-driven current, exhibiting multi-mode chirp dynamics; (b) Mode jumping phenomenon of QCL1 under a 186 mA pulse-driven current. All measurements were performed at a fixed heat sink temperature of 11.3 K, with CW-driven QCL2 biased at 230 mA and operated as the LO. The IF signals are generated by multi-heterodyne beating between QCL1 (pulsed) and QCL2 (LO). The system bandwidth is ~3 GHz (limited by the amplifier); the resolution bandwidth is 10 MHz, determined by the 0.1 µs STFT slice duration.} 
\label{f6}
\end{figure}

In Fig. \ref{f1}(b), we show that for CW-driven operation, QCL1 transitions from a single-mode regime to a multi-mode regime when the driving current exceeds 208 mA. The single-mode chirp dynamics of QCL1 under selected pulsed current conditions were investigated in Fig. \ref{f3}, leading us to infer that a further increase in pulsed current would enable the detection of QCL1’s multi-mode chirp dynamics. Experimental evidence for this multi-mode behavior is provided in Fig. \ref{f6}, where three down-chirped frequency components are detected in the RF band.
The multi-mode chirp dynamics shown in Fig. 6 arise from a multi-heterodyne beating process between QCL1 and QCL2. In this measurement, QCL2 operates as the LO with a lasing frequency of approximately 4.2 THz. At the bias current of 230 mA, QCL2 operates in a three-mode state. The three chirped modes of QCL1 beat with the nearest-neighbor modes of the LO, generating three down-converted IF signals in the RF band, as shown in Fig. 6.
When QCL1 is driven by an 180 mA pulsed current, as shown in Fig. \ref{f6}(a), three chirped curves that with an approximate spacing of 300 MHz between adjacent curves exhibit a consistent down-chirp behavior over the 1000 $\mu$s pulse duration. The gray shaded region spanning from 150 $\mu$s to 350 $\mu$s indicates that the three modes in the THz region possess different chirp rates. During this 200 $\mu$s interval, the down-chirp ranges of the three modes are 60 MHz, 40 MHz and 110 MHz, respectively. We also observe that mode 3 exhibits a continuous, complete chirp curve throughout the 1000 $\mu$s duration, while mode 2 appears as discrete data points in some segments. This is because mode 3 has a stronger signal intensity, whereas the SNR of mode 2 is relatively poor. The spectral slices in Fig. S6 of the Supplementary Material demonstrate this intensity difference---this prevented reliable identification of the weak signal peak position during these periods, leading to incomplete data acquisition. We therefore deduce that the two lasing modes with an approximate spacing of 600 MHz in the multi-mode regime (Fig. \ref{f1}(b)) correspond to mode 1 and mode 3 in Fig. \ref{f6}(a). Beyond these two modes, there should be an additional intermediate lasing mode corresponding to mode 2. However, this intermediate mode was not resolved in Fig. \ref{f1}(b) due to the insufficient amplitude resolution of the spectrum analyzer.
When the pulsed current is further increased to 186 mA , mode jumping\cite{khalatpour2021high, wysocki2005widely} occurs during the chirp process of the three modes. Fig. \ref{f6}(b) illustrates this phenomenon: at approximately 60 $\mu$s, abrupt frequency shifts occur for all three modes. Specifically, over the time window of 35 $\mu$s to 85 $\mu$s, modes 1 and 3 exhibit a 100 MHz downward frequency shift, while mode 2 undergoes a 40 MHz upward frequency shift.
The abrupt frequency shifts observed in Fig. 6(b) at approximately 60 $\mu$s are referred to as “mode jumping”—small scale frequency jumps occurring within individual longitudinal modes—rather than full longitudinal mode hopping between adjacent longitudinal modes (which would correspond to the ~41.6 GHz free spectral range of the 1 mm cavity).
Spectral slices associated with the relevant time points are illustrated in Fig. S7 of the Supplementary Material. By the end of the 1000 $\mu$s pulse duration, the spacing between adjacent curves returns to approximately 300 MHz, which is consistent with the spacing observed in Fig. \ref{f6}(a). The observation of this abrupt mode jumping demonstrates the high-speed performance of the system, a property that is crucial for potential transient spectroscopy detection applications.

\section{Discussion}           
In this work, we report a systematic study on the transient chirp dynamics of THz QCLs using an on-chip heterodyne detection scheme, and explore their physical mechanisms and practical application potential. The highlights of our work lie in the following aspects. Firstly, our work fills the research gap in the study of chirp in THz QCLs. Notably, previous studies on transient chirp dynamics in THz QCLs are mostly on sub-$\mu$s timescales, with rate-equation-related theory or Maxwell--Bloch equations to model their transient carrier--photon dynamics. Yet they have not addressed the $\mu$s--ms thermal chirp when thermal accumulation rather than carrier-photon coherence drives the frequency evolution. The few existing thermal chirp investigations are also focused on the mid-infrared band, where mature, high-performance detectors with room-temperature operation are available. This has made such research relatively straightforward to conduct and enabled the observation of chirp bands spanning tens of gigahertz (GHz). In contrast, the THz band suffers from a relative lack of efficient, practical detectors.
 Here, by employing THz QCLs, an ideal dual-functional device serving as both the radiation source and detector for the THz band, we have conducted preliminary investigations into frequency chirp dynamics in the THz band. 
Secondly, our on-chip heterodyne detection scheme provides a simple yet effective measurement capability for thermal chirp investigations. The observation and classification of three distinct chirp profiles is demonstrated here for the first time.
Thirdly, our work provides valuable insights for the practical application of frequency chirp in the THz band. Although the chirp span in the THz region is much narrower compared to that in the mid-infrared region, and the chirp range of THz QCLs under single-pulse pumping conditions is limited to the hundreds-of-MHz order of magnitude, they still hold considerable potential for diverse applications. As illustrated in Fig. \ref{f3}, the intrinsic frequency-sweeping characteristic of the chirp behavior enables us to achieve gapless frequency coverage across the 810–1550 MHz range by tuning only 8 sets of currents. This capability can be exploited for broadband THz absorption spectroscopy of molecules, significantly enhancing the efficiency of material identification. As shown in Fig. \ref{f1} (b), the linear frequency-current relationship in the single-mode regime of QCLs enables frequency-modulated (FM) communication, where information can be flexibly encoded into parameters such as current amplitude and pulse width, and subsequently decoded from chirp frequency data. Third, frequency chirp can be regarded as an intrinsic response to strong current perturbation. In-depth investigation of such chirp behavior not only facilitates a deeper understanding of the internal nonlinear dynamic processes in THz QCLs but also provides critical physical insights for advancing the flexible tuning of THz QCL lasing frequency and the development of current-locked THz QCL optical frequency comb technology.

Despite these promising aspects, several limitations of our current implementation should be acknowledged.
 While our on-chip twin-QCL architecture serves as a compact and effective platform for investigating THz QCL chirp dynamics, it is not yet directly adaptable for practical applications, owing to the absence of efficient free-space optical coupling. Moreover, the shared substrate of the two QCLs results in inadequate electrical and thermal isolation between them, further limiting their immediate practical deployment. 
 In our simulations, the physical model is simplified and only accounts for the thermal effect of electrical current on the frequency chirp dynamics of THz QCLs. This model can provide a basic explanation for the overall trend of chirp dynamics on $\mu$s--ms timescales. However, the carrier--photon dynamics on sub-$\mu$s timescales are not included and unable to be explained by our current model. And it also cannot explain the emergence of multimode sidebands or the underlying mechanism of mode jumping when the current increases.
 
It is worth noting that to further improve the simulation model, the thermal model can be extended by coupling it with carrier--photon rate equations to capture the sub-$\mu$s transients and cover the full ps--ms timescale chirp dynamics. Furthermore, for the multi-mode chirp and mode-jumping phenomena observed in Figure 6, Maxwell--Bloch equations, which naturally account for multimode coupling, can provide a unified framework capable of elucidating the underlying mechanisms.
 
\section{Conclusion}
In summary, we have systematically investigated the transient thermal chirp dynamics of pulse-driven THz QCLs using an on-chip heterodyne detection method. By extracting the chirp frequency peaks from the spectrogram of sliced time-domain signals, we experimentally characterized the chirp behaviors of QCLs in both single-mode and multi-mode regimes, achieving a temporal resolution of 0.1 $\mu$s and a frequency resolution of 10 MHz. By integrating driving current tuning with frequency chirp, we achieved gapless frequency coverage spanning 810 MHz to 1550 MHz. This current-tuning capability also allowed flexible adjustment of the relative lasing frequency positions between the two QCLs, thereby revealing three distinct down-converted chirp profiles. Furthermore, the two-node thermal model we developed effectively accounts for the chirp reversal phenomenon observed in experiments, laying a solid foundation for a more theoretical and in-depth understanding of the frequency chirp dynamics of THz QCLs. Collectively, these findings not only deepen the fundamental understanding of THz QCL chirp behaviors but also provide critical guidance for developing advanced functionalities of THz QCLs, underscoring the significant scientific and application value of our research.

\section{Experimental Section}
\threesubsection{THz QCL fabrication} The THz QCLs employed in this work feature a hybrid active region design, which leverages bound-to-continuum transitions for THz photon emission and fast longitudinal optical phonon scattering to efficiently depopulate the lower laser state and ensure population inversion. Tailored for light emission at 4.2 THz, the QCL active region employs a detailed layer structure as described in prior work\cite{wan2017homogeneous}. The entire active region structure was grown on a semi-insulating GaAs (100) substrate via a molecular beam epitaxy (MBE) system, and the grown wafer was subsequently processed into a single plasmon waveguide geometry using traditional laser micro-fabrication technologies, with each waveguide adopting an optimal ridge width of 150 $\mu$m. To enhance thermal management, the laser substrate was thinned to 100 $\mu$m via grinding and polishing, after which the processed laser ridges were cleaved into laser bars with a nominal cavity length of 1 mm. The cleaved laser bars were then patterned into an on-chip twin-device configuration, with a 1.6 mm spacing between the two devices. Finally, the twin-device laser bars were indium-bonded onto copper heat sinks for subsequent wire bonding and device testing.

\threesubsection{Chirp frequency extraction} To investigate the transient chirp dynamics of the THz QCL, the down-converted heterodyne signals were captured using a high-speed real-time oscilloscope (Teledyne LeCroy, Wave-Master 820 Zi-B). After applying a digital band-pass filter to isolate the beat-note signal, the data was partitioned into consecutive 0.1 $\mu$s temporal slices. We utilized a Short-Time Fourier Transform (STFT) approach to generate a time-evolving spectrogram, where spectral peaks were extracted to form a candidate frequency matrix. Due to the varying chirp rates—which are significantly higher during the initial current injection phase compared to the steady-state emission—we implemented a robust, automated tracking pipeline. This method utilizes a bidirectional traversal strategy starting from a stable reference point to reconstruct the frequency trajectory. To maintain high fidelity across different regimes, the algorithm employs zone-adaptive frequency thresholds: a wider search window accommodates the rapid transient phase, while a narrower window suppresses noise in the stable region. Furthermore, a look-ahead mechanism was integrated to ensure trajectory continuity in instances of transient signal fading by interpolating across skipped time-slices (see Algorithm S1 in the Supplementary Material for the detailed logical implementation).

\medskip
\textbf{Supporting Information} \par %Please delete the Suppporting Information statement if it is not applicable. Please supply Supporting Information in another file. Supporting information should not be provided in .tex format
Supporting Information is available from the Wiley Online Library or from the author.

% Acknowledgements
\medskip
\textbf{Acknowledgements} \par %delete if not applicable))
X. Bi and X. Ma contributed equally to this work. This work is supported by the National Key Research and Development Program of China (2025YFE0217500), the National Natural Science Foundation of China (62325509, 62235019, 62505344, 62531005, 62575299, 62275258, 62305364, 62435017, T2550072 and 12333012), Science and Technology Commission of Shanghai Municipality (23ZR1474000), the CAS Project for Young Scientists in Basic Research (YSBR-069), and Autonomous deployment project of State Key Laboratory of Materials for Integrated Circuits (SKLJC-Z2025-B03), Youth Innovation Promotion Association, Chinese Academy of Sciences (2023241). The authors gratefully acknowledge the technical and equipment support  provided  by ShanghaiTech Material and Device Lab (SMDL).

\textbf{Conflict of Interest} \par
The authors declare no conflict of interest.

\textbf{Data Availability Statement} \par
The data that support the findings of this study are available from the corresponding author upon reasonable request.
% References
\medskip

% Use the following code if you wish to generate your bibliography with BibTeX;
% replace the string "MSP-template" below with the name(s) of
% the BibTeX data base(s) you want to use.
% The resulting bibliography-output (the content of the .bbl file)
% must be pasted back into this file before submission.
% Please also include your BibTeX data base file(s) in your submission
% so that we can re-run BibTeX if necessary.
%

\bibliographystyle{MSP}
\bibliography{Reference}

\end{document}